\documentclass[oupdraft]{bio}
\usepackage{url}

\usepackage{mathtools}
\usepackage{amsmath}
\usepackage{amsmath,amssymb,amsfonts,bm}
\usepackage[outer=1.8cm, inner=1.8cm]{geometry}

\usepackage{algorithm,algorithmicx,algpseudocode,setspace}
\usepackage{booktabs}
\begin{document}

\title[Generic Sleep-Wake Cycle Detection]{Validating CircaCP: a Generic Sleep-Wake Cycle Detection Algorithm}

\author{Shanshan Chen$^\ast$, Xinxin Sun\\[4pt]
\textit{Department of Biostatistics, School of Medicine, Virginia Commonwealth University}
\\[2pt]
{shanshan.chen@vcuhealth.org}}

\markboth%
{Shanshan Chen and Xinxin Sun}
{Generic Sleep-Wake Detection Algorithm}

\maketitle

\footnotetext{To whom correspondence should be addressed.}

\begin{abstract}
{Sleep-wake cycle detection is a key step when extrapolating sleep patterns from actigraphy data. Numerous supervised detection algorithms have been developed with parameters estimated from and optimized for a particular dataset, yet their generalizability from sensor to sensor or study to study is unknown. In this paper, we propose and validate an unsupervised algorithm -- CircaCP -- to detect sleep-wake cycles from minute-by-minute actigraphy data. It first uses a robust cosinor model to estimate circadian rhythm, then searches for a single change point (CP) within each cycle. We used CircaCP to estimate sleep/wake onset times (S/WOTs) from 2125 indviduals' data in the MESA Sleep study and compared the estimated S/WOTs against self-reported S/WOT event markers. Lastly, we quantified the biases between estimated and self-reported S/WOTs, as well as variation in S/WOTs contributed by the two methods, using linear mixed-effects models and variance component analysis. \\
On average, SOTs estimated by CircaCP were five minutes behind those reported by event markers, and WOTs estimated by CircaCP were less than one minute behind those reported by markers. These differences accounted for less than 0.2\% variability in SOTs and in WOTs, taking into account other sources of between-subject variations. By focusing on the commonality in human circadian rhythms captured by actigraphy, our algorithm transferred seamlessly from hip-worn ActiGraph data collected from children in our previous study to wrist-worn Actiwatch data collected from adults.  The large between- and within-subject variability highlights the need for estimating individual-level S/WOTs when conducting actigraphy research. The generalizability of our algorithm also suggests that it could be widely applied to actigraphy data collected by other wearable sensors.}\\
{Actigraphy, sleep-wake cycle detection, parametric change point detection, MESA Sleep, event markers,  validation}
\end{abstract}\\

\newpage
\section{Introduction}
\label{sec1}
Accurate sleep-wake (SW) cycle detection is essential for extrapolating sleep patterns from actigraphy data (\cite{Meltzer2012}). Actigraphy records one's activity levels at 30-second or 60-second intervals continuously for a prolonged time (e.g. a few days to several weeks). Using such records, researchers can estimate circadian cycles and SW cycles, and extrapolate sleep metrics such as sleep/wake onset times (S/WOTs), sleep durations, and the within-subject variability of these sleep metrics. \par  

Numerous algorithms for estimating SW cycles from actigraphy data have been developed and such research attempts date back to the 1980s (\cite{Mullaney1980} \cite{Webster1982}). Mullaney et al were the first group to devise such algorithms and then Webster et al proposed a scoring algorithm (\cite{Webster1982}). Other researchers extended Webster's scoring method and devised their own scoring algorithm customized to their data (\cite{Cole1992} \cite{Sadeh1994},\cite{Oakley1997}). These scoring algorithms have been widely used to identify SW states due to their accessibility and ease of use. In recent years, machine-learning techniques have also been applied in S/WOT detection. Most of these techniques are binary classification algorithms such as logistic regression \cite{Sazonov2004}\cite{Paquet2007}\cite{Domingues2013}, linear discriminant classifiers (\cite{Long2017}), support vector machines (\cite{Palotti2019}), and random forests (\cite{Angelova2020} \cite{El2017}). \par

These algorithms have several disadvantages. First, algorithms based on generalized linear models or supervised machine-learning algorithms require ground-truth labels to train models for SW cycle detection. Labels for training models can be expensive to collect (e.g. from polysomnography (PSG) studies), contain inaccuracies (e.g. proxy reports), or be inconvenient to collect (e.g. researchers may elect not to collect labels in order to maximize the compliance rate). Secondly, since the debut of actigraphy sensors in the 1980s, activity levels have been measured in relative scales without a consistent unit. Thus, actigraphy data collected by different wearable activity trackers for the same types of activity have different ranges and distributions. As the parameters of these algorithms are fine-tuned to maximize the detection accuracy for a particular dataset, their generalizability to other datasets is questionable. Thirdly, these algorithms use fixed-length time windows to extract features, which cannot adequately capture the structural breaks in data distributions that may happen on various timescales (\cite{Auger1989}). Moreover, these algorithms assumed that the distributions of diurnal and nocturnal actigraphy are identical. As a result, the SW cycles estimated by these algorithms tend to have multiple mis-identified sleep/wake states, resulting in multiple fragments of sleep/awake states.  These estimated SW cycles may be further smoothed by an ad hoc temporal filter (e.g. a median filter) (\cite{Cao2012}) or heuristic rules (\cite{Webster1982}). Although such filters may consolidate some of the fragments and improve detection accuracy (\cite{Palotti2019}), they can also further reduce the temporal resolution of the detection. Figure 1 illustrates this common issue, based on the application of three of these scoring algorithms to the actigraphy data of one subject. \par
Recognizing these issues, some researchers have devised algorithms focusing on general SW cycle detection. Van Hees et al descried a accelerometer-based estimation method without sleep labels; instead, heuristic cutoffs were used to estimate various postures and corresponding SW states (\cite{VanHees2018}). Cakmak et al fused wrist-worn sensor data collected at higher sampling rates, applied a binary segmentation algorithm to identify the change points, and integrated three types of segmented time series by training generalized linear models (\cite{Cakmak2020}). These algorithms were designed in particular for wearable sensor data stored at higher temporal resolutions (e.g. 10Hz), and cannot be readily applied to actigraphy data that are aggregated at 30-second or 1-minute intervals. \par

Chen et al (\cite{Chen2019}) proposed a generic, unsupervised algorithm to detect sleep-wake cycles and estimate S/WOTs based on common characteristics in human circadian rhythms. The algorithm leverages the prior information obtained from a nonlinear parametric model, and detects the S/WOTs with higher precision using a parametric change point (CP) detection algorithm. This algorithm was tested on 112 children's data collected from hip-worn ActiGraph\textsuperscript{TM} sensors without ground-truth labels, and its generalizability has not yet been tested. In this paper, we detail the mathematical foundation of Chen's algorithm (\cite{Chen2019}), improve its computation time and robustness, and validate the algorithm on a large actigraphy dataset collected by wrist-worn Actiwatch sensors used in the Multi-Ethnic Study of Atherosclerosis (MESA) Sleep study (\cite{Zhang2018}\cite{Chen2015}). This dataset contains week-long actigraphy data from over 2000 adults of diverse ages, ethnicities and work schedules, and are accompanied with self-reported S/WOTs recorded as event markers, making them perfect for external validation. \par

\section{Methods}\label{Methods}

\subsection{Data Collection}
A total of 2237 subjects aged from 45 to 84 were enrolled in the MESA sleep study between 2010 and 2012. During the study, all 2237 subjects were instructed to place an Actiwatch (Philips Respironics, Inc) on the non-dominant wrist, and wear the same sensor for at least five days in the habitual living environment. Subjects were also asked to press the small button located on the right side of the Actiwatch every time they went to sleep or woke up, creating event markers. After the study days, the devices were returned to the study staff, and the data from these devices were downloaded by Respironics Actiware 5 Software, and aggregated as activity counts in 30-second intervals using the Actiware-Sleep software. The quality of these data was scored by the Sleep Reading Center at the Brigham and Women's Hospital. 

\subsection{Measures}
A total of 2159 subjects in the MESA Sleep study had actigraphy data, of whom 2125 had a minimum of three days of data that was no less than 50\% reliable. These 2125 subjects entered our actigraphy analysis. The quality of these actigraphy data were rated from 2 to 7 (2 being the poorest and 7 being the best, named as G2, G3,..,G7 in the rest of this paper) according to consistency with PSG data, event markers, sleep diary and light levels (more details can be found in the MESA Exam 5 Sleep Data Documention Guide from sleepdata.org/datasets/mesa). Note that quality grades do not necessarily reflect the quality of the actigraphy data alone, but also depend on the completeness and quality of sleep diaries. We used a one-dimensional time series of the actigraphy data --- the vector magnitude of activity counts --- for SW cycle detection, and compared the detection results with the event marker data to validate the SW cycles detected by our algorithm. \par

\subsection{Screening of Actigraphy Data}
To reliably estimate circadian rhythms, we included subjects who had worn the sensor continuously for at least four days. To do so, we first excluded subjects with actigraphy of less than 5,760 minutes. Then among the remaining subjects, we detected the continuous wearing periods, defined as having no consecutive zeros more than 120 minutes. We included subjects with a longest continuous wearing period of at least 5,760-minutes. Lastly, we aggregated the included 30-second interval actigraphy into 60-second interval actigraphy.  

\subsection{Sleep Detection Algorithm}
This section details Chen's algorithm for SW cycle detection. The algorithm consists of two steps, 1) segmenting the time series of actigraphy data into circadian cycles using a non-linear parametric model; each cycle consists of a diurnal and a nocturnal period, and 2) identifying more accurate sleep-to-wake and wake-to-sleep CPs within each segment. \par
The time series of human activity data demonstrate a strong pseudo-periodical pattern (i.e. circadian rhythm), which can thus be modeled by functions with periodical patterns, e.g. a cosine function. Halberg et al (\cite{Halberg1967})  proposed a cosinor model to fit data with circadian patterns. It has three parameters, amplitude $amp$ which is the peak of the rhythm, $mes$ which indicates the midline estimating statistic of rhythm, and acrophase $phi$ which is the time of the peak of the rhythm, to be estimated to identify the circadian rhythm. The cosinor model is shown in Equation 1:
\begin{align}
	r(t) = mes + amp \times cos(\dfrac{[t-\phi]\times2\pi}{T})
\end{align}
where t = 1, 2, ..., n, t $\in\mathbb{N}$, is the time index for the actigraphy sequence.  T is the period of one's circadian rhythm, often set as a constant of 24 (hours), or in our case 1440 (minutes). Marler's  extension of this model, the sigmoidally-transformed cosine model, can also be used to capture the circadian rhythm. Since the cosinor model takes significantly less computation time without sacrificing much accuracy, our study implemented the cosinor model to fit the non-linear curve. We estimated the parameters of the non-linear curve by minimizing the sum of squared residuals. The objective function we optimized is displayed in the formula below:

\begin{align}
	\hat{\mathbf{p}} = \underset{\mathbf{p} \in\mathbb{R}^{3} }{argmin}\ \lVert {F(\mathbf{p},t)-Y}\rVert
\end{align}

where $\mathbf{p}$ is a 3-parameter vector containing $mes$, $amp$ and $\phi$ in Equation 1, the initial value of $\mathbf{p}$ is set as (500, 550, 227). $F(\mathbf{p},t)$ is the fitted curve and $Y$ the raw activity counts with sequence index $t$, $ t\in \mathbb{N}$. After the non-linear curve fitting, we dichotomize the fitted curve $F$ by treating the lower 18\% of its range as nocturnal periods and the upper 82\% as diurnal periods. This dichotomized curve gives an estimate of one's circadian cycles whose transition edges are rough estimates of S/WOTs.

The fixed period of the cosinor model imposes consistent S/WOT over the days and hence cannot capture day-to-day variation in S/WOT for each individual. Therefore, the precise sleep/wake times must be determined with a more refined searching algorithm. Next, we locate more precise S/WOTs for a particular circadian cycle identified by the cosinor model. If a subject's actigraphy sequence has clear circadian patterns, we assume the subject gets up at least once between two consecutive SOTs, or goes to sleep at least once between two consecutive WOTs, during a day. As a subject's activity pattern shifts most drastically from night to day or day to night, we can assume that S/WOTs are the most significant structural break points before and after which the data distributions were drastically different. Such structural CPs can be detected using a parametric change point detection method. 

Although count data are traditionally modeled by Poisson distributions, activity "counts" are not natural counts that are generated from a counting process, but intensity measures of activity based on various algorithms. Distributions of such intensity data (e.g. time series of seismic moments and precipitation intensity) usually show long-tail, zero-inflated patterns, and have been modeled by Gamma distributions (\cite{Kagan2002}\cite{Martinez2019}). Thus, we fit Gamma distributions to MESA actigraphy data, as shown in Figure 2. Activity distributions during waking and sleep are distinctively different: the activity counts during sleep have a higher density at 0 compared to those during waking periods.

Thus, we describe the time series of actigraphy as a pseudo-cyclostationary random process Y, which collects a sequence of independent and identically distributed random variables, $Y=[y_{1},y_{2},...,y_{k},...,y_{n}]$, that follow Gamma distributions, i.e.,
\begin{align}
	y_{k}\sim Gamma(\theta_{k},\xi)
\end{align}

where $\theta_{k}$ is the scale parameter for the random variable at time $k$, $y_{k}$, and $\xi$ the shape parameter shared by the random variables $[y_{1},y_{2},...,y_{k},...,y_{n}]$. A structural change point in such a time series means the scale parameter of the sequence before and after this point is significantly different. To test if such a structural change point exists in a segment of time series data, we perform a test based on likelihood ratios. The hypothesis can be expressed by the following mathematical notation:
\begin{align}
	H_{0}: \theta_{1} = \theta_{2}= ...= \theta_{k} =... = \theta_{n}=\theta_{0} \\
	H_{1}: \theta_{0}=\theta_{1} = \theta_{2}=...=\theta_{k} \neq \theta_{k+1} =...=\theta_{n}  
\end{align}
where $k$ is the location of the change point, $n$ the length of the data sequence, and $\delta$ a real number. Under $H_{0}$, the likelihood function is:
\begin{align}
	L_{0}=\dfrac {\prod_{i=1}^{n} y_{i}^{\xi-1}} {\Gamma^{n}(\xi)} \dfrac{1}{\theta_{0}^{n\xi}} \exp\dfrac{-\sum_{i=1}^{n}y_{i}}{\theta_{0}}
\end{align}

whereas under $H_{1}$, the likelihood function is
\begin{equation}
	\begin{multlined}
		L_{1}(\theta_{1},\theta_{2})= \dfrac {\prod_{i=1}^{n} y_{i}^{\xi-1}}{\Gamma^{n}(\xi)} \dfrac{1}{\theta_{1}^{k\xi}}
		\dfrac{1}{\theta_{2}^{(n-k)\xi}} \exp \dfrac{-\sum_{i=1}^{k} y_{i}}{\theta_{1}} \cdot \exp \dfrac{-\sum_{i=k+1}^{n} y_{i}}{\theta_{2}}
	\end{multlined}
\end{equation}

where $j = 1,2,...,k,...,n-1$, and represents all the possible locations of a change point $k$. Thus, to estimate the location of the change point $k$ in a time series, our goal is to find the maximum likelihood ratio $\dfrac{L_{1}}{L_{0}}$,  or the minimum of the Bayesian Information Criteria (BIC). Chen et al \cite{Chen2006} proposed an improved criteria, called modified information criteria (MIC), which is particularly suited for assessing model complexity in change-point models. Thus, here we adapted the MIC term with a scaling factor $\lambda$, $\lambda$ controls the penalty of the change point being detected near the beginning or near the end of a sequence:
\begin{equation}\label{eqbic}
	\begin{multlined}
		MIC(k) = -2 \log(L_{1}(\hat{\theta_{1}},\hat{\theta_{2}}))+2\log n\\
		= 2k\xi \log\sum_{i=1}^{k} y_{i} + 2(n-k)\xi  \log\sum_{i=k+1}^{n} y_{i} \\
		 -2k \log k\xi -2(n-k) \log(n-k)\xi \\
		 - 2(\xi-1)\sum_{i=1}^{n}\log y_{i} + 2\log n +\lambda\times(\dfrac{2k}{n}-1)^2 \log n
	\end{multlined}
\end{equation}

Since the third and fifth terms in the equation of $MIC(k)$ do not contain location parameter $k$, we dropped the fifth and sixth terms from the model and simplified the fourth term. Therefore, Equation \ref{eqbic} reduces to:
\begin{equation}\label{eqk}
	\begin{multlined}
		\hat{k} = \underset{k \in\mathbb{N}}{argmin}\ [ 2k\xi \log\sum_{i=1}^{k} y_{i} + 2(n-k)\xi \log \sum_{i=k+1}^{n} y_{i}- 
		2k\xi \log(n\xi)-2(n-k)\xi \log((n-k)\xi) + \lambda\times(\dfrac{2k}{n}-1)^2 \log n]
	\end{multlined}
\end{equation}
$\hat{k}$ is the estimated location of the change point in sequence $Y$. $\lambda$ is empirically determined as 50. The whole algorithm for sleep-wake cycle detection is detailed in the Algorithm \ref{algo1}. A Matlab implementation of CircaCP can be found at https://github.com/ShanshanChen-Biostat/CircaCP.

\begin{algorithm}\label{algo1}
	\small
	\caption{Detecting precise S/WOTs guided by the cosinor model}\label{algo1}
	\begin{algorithmic}[1]	

		\Require $Y$, $Y=[Y_{1},Y_{2},...,Y_{n}]$ \Comment{Actigraphy sequence of a subject}
		\Ensure $Y$ is positive, e.g.  $ Y= Y+1e-1$ \Comment{to fit Gamma distributions}
		\State  Fit the cosinor model in Equation 1 to $Y$ and get the estimated curve $F$
		\State  Dichotomize $F$ to get a binary sequence $D$:
		$$
		\begin{cases}
			D_{i} = 1, \quad when\quad F_{i}>0.18 \times range(F) \\
			D_{i} = 0, \quad when\quad F_{i}<=0.82\times range(F)
		\end{cases} 
		$$
		\State  $E_{i} = D_{i+1}-D_{i}$, $E= [E_{1},E_{2},...,E_{n-1}]$ \Comment{first-order, finite difference of $D$}
		\State  $B = index(E\neq 0)$, $B=[B_{1},B_{2},..,B_{m}]$ \Comment{rough boundaries of sleep-wake cycles detected by the cosinor model}
		\State  $EWT = index(E=1)$, \Comment{index of roughly-identified WOTs}
		
		\If {$length([Y_{1},Y_{2},...,Y_{B_{2}}]) >240$}  \Comment{detect change point if there are enough samples in the beginning}
		\State search for the first single change point $CP_{1}$ in $[Y_{1},Y_{2},...,Y_{B_{2}}]$
		\Else
		\State $CP_{1} = B_{1}$
		\EndIf
		\If {$B_{2} \in EWT$}  \Comment{determine whether the first change point is SOT or WOT}
		\State $Label_{CP_{1}} = SOT$
		\Else
		\State $Label_{CP_{1}} = WOT$
		\EndIf
		
		\For {each $B_{i}$ in $[B_{2},B_{2},..., B_{m-2}]$}
		
		\State $[\theta,\xi] = mle(X,Gamma)$, \Comment{Maximum likelihood estimates of $X \sim Gamma(\theta,\xi)$} 
		
		\For {each $k$ in $[1:l]$}
		\State $X = [Y_{CP_{i-1}},...,Y_{B_{i+1}}]$, length of $l$ \Comment{the segment between the previous precise CP and the next possible CP}
		
		\If {$B_{i+1} \in EWT$} \Comment{determine whether the current change point is SOT or WOT}
		\State $Label_{CP_{i}} = SOT$; \Comment{label the detected current change point as SOT}
		\Else
		\State $Label_{CP_{i}} = WOT $; \Comment{label the detected current change point as WOT}
		\EndIf
		
		\State $Seg_{1} = [X_{1},X_{2},...X_{k}] $  \Comment{the first half of the sequence before $k$}
		\State $Seg_{2} = [X_{k+1},X_{k+2},...X_{l}] $  \Comment{the second half of the sequence after $k$}
		\State calculate $MIC_{k}$ in Equation \ref{eqk} 
		\EndFor
		\State $loc_{min} = {argmin}(MIC)$ \Comment{the index of the minimum value on the MIC curve}
		\State $CP_{i} = loc_{min} + CP_{i-1}$ \Comment{match global index}
		\EndFor
		
		\If {$length([Y_{B_{m}},...,Y_{n}]) >240$} \Comment{detect change point if there are enough samples in the end}
		\State search for the last single change point $CP_{m}$ in $[Y_{B_{m}},...,Y_{n}]$
		\Else
		\State $CP_{m} = B_{m}$
		\EndIf
		
		\If {$CP_{m-1} \in EWT$}   \Comment{determine whether the last change point is SOT or WOT}
		\State $Label_{CP_{m}} = SOT$
		\Else
		\State $Label_{CP_{m}} = WOT$
		\EndIf
		\State Repeat Step 6 to Step 40, get more precise estimates
		
		\State \textbf{Output:} a sequence of precise change points $CP = [CP_{1},CP_{2},...,CP_{m}]$ labeled as SOT or WOT
	
	\end{algorithmic}
\end{algorithm}

\subsection{Error Detection}
To identify detection errors without ground-truth labels, we adopted a metric for evaluating clustering performance -- the Calinski-Harabasz (CH) criterion (\cite{Calinski1974}). Also known as the variance ratio criterion, this metric measures the ratio of total between-cluster sum of squares (SSB) to total within-cluster sum of squares (SSW). 

\begin{align}
	CH = \dfrac{SSB}{\sum_{i=1}^{k}SSW_{k}} \times \dfrac{n-k}{k-1}
\end{align}

where $n$ is the total number of samples in an actigraphy sequence, $k$ is the number of clusters, in our case, $k=2$ (i.e. sleep and awake periods). We evaluated both the $CH$ for SW cycles estimated by the cosinor model ($CH_{cos}$) and by our proposed algorithm ($CH_{pro}$). Detection results with $CH_{pro} -CH_{cos} <100$ were identified as containing detection errors and removed from the subsequent analysis.

\subsection{Validation Procedure}

Self-reported event markers contain inherent errors. For example, the subject may forget to click the button before falling asleep or after waking up, or click the button multiple times for a true S/WOT, or even at a random time of day which was not bed time or wake-up time. Thus, for validation, we paired up the S/WOTs detected by the CircaCP algorithm with the most likely corresponding event markers. First, sequence alignment methods such as the Needleman-Wunsch algorithm (\cite{Needleman1970}) and Dynamic Time Warping (\cite{Mueen2016}) were applied to match event markers and the estimated S/WOTs. However, these existing alignment methods did not work well because long gaps between event markers and S/WOTs are quite common when the subject forgets to record S/WOTs. On the other hand, the correct event markers can always be detected close to true S/WOTs. Hence, we searched for event markers within the proximity of the estimated S/WOTs directly. Here, we defined a marker as valid if it fell within 180 minutes of the estimated S/WOT. If multiple markers matched for a certain S/WOT, we kept only the latest marker for SOT and the earliest marker for WOT. 

Once the event markers and the estimated S/WOTs were matched, we converted the index-based S/WOTs and event markers to minute-based S/WOTs, defined as the time elapsed since midnight, for quantitative analysis. SOTs occurring after midnight had 1440 minutes added to the original elapsed time.

To assess the variability in S/WOTs from the two sources (i.e. event markers and CircaCP estimation) and from actigraphy of different quality, we conducted variance component analyses to decompose the total variance in the data into the percentage contributions of various random effects (i.e. algorithm, subject and within-subject, day-to-day variability).  We also assessed the effects of these factors on the outcome --- SOT or WOT --- using linear mixed-effects models. Specifically, actigraphy quality group, algorithm, work schedule, age, gender and ethnicity were modeled as fixed effects. The reference levels were the highest actigraphy quality group (i.e. G7), event markers, being white, female, and youngest, and having a day job. Day-to-day variability was modeled as random effects nested within subjects. Lastly, we assessed the agreement between the two methods using Bland-Altman analysis. 

\section{Results}\label{sec5}

Out of the 2125 subjects' actigraphy data, 1957 subjects' actigraphy passed the screening criteria (i.e. having a continuous wearing time of at least four days) and were used for SW detection (Table \ref{tab1}). Our screening criteria kept more than 90\% of subjects' actigraphy of varying quality, with quality group G7 having the most subjects included. Among these 1957 subjects, 34 subjects with poor SW detection results (evaluated by the error detection procedure) were excluded. Of the remaining 1923 subjects, 1857 had at least one matched marker and algorithm-detected S/WOT pair and entered the validation analysis. The demographics of these 1857 subjects are also listed in Table \ref{tab1}. The majority of these subjects were retired, aged between 66 and 71, white, and female. In the group with the lowest actigraphy quality (G2), subjects were older and mostly African-American, and had a higher percentage of shift jobs than other groups. The G2 group had the lowest percentage (88.9\%) of actigraphy that entered the validation analysis whereas in the G4 to G7 group, almost all actigraphy sequences for SW detection entered the validation analysis. \par     

The SW cycle detection results are visualized in Figure 3. Despite the variety of work schedules, the CircaCP algorithm correctly identified SW cycles, where the transition edges are the estimated S/WOTs. These S/WOTs are very close to the S/WOTs reported by the event markers. Bland-Altman plots in Figures 4 and 5 show the limits of agreement between the estimated S/WOTs and those recorded as the event markers. Although our marker matching procedure implied that the maximum difference between matched markers and estimated S/WOTs is $\pm$180 minutes, the limits of agreement between the two were well under 180 minutes (about 100 minutes for SOT and 90 minutes for WOT). Our variance-component model further analyzes the factors that impact the S/WOT outcomes (Table \ref{varcomp}). Overall, the factor of S/WOT estimation method contributes less than 0.2\% to the total S/WOTs variance, suggesting that the CircaCP algorithm is equivalent to the event markers when analyzing the impact of the estimation algorithm on the outcomes. The majority of the variation comes from the subject-to-subject variability and day-to-day variability.\par

Results from linear mixed-effects models are shown in Table \ref{mixed}. SOTs estimated by our proposed algorithm are about 6 minutes behind SOTs reported by markers in the highest-quality actigraphy group, and more behind (though non-significantly) in worse-quality actigraphy groups. WOTs detected by CircaCP are almost identical to those recorded by markers, with a non-significant difference of less than one minute. On average, the subjects who are the youngest white females with a day job, and have the best actigraphy quality slept at 23:23 and woke up at 6:45. Having a non-regular work schedule significantly affects the S/WOT outcomes, and working night shifts delays WOTs the most among all types of schedule. Among the four race groups, African Americans had significantly delayed SOTs in comparison to white Caucasians. Since the quality grades do not entirely reflect raw actigraphy quality, this factor (i.e. the quality groups) did not impact the S/WOT outcomes significantly. \par

\section{Discussion}
In this study, we improved and validated a generic, unsupervised algorithm for detecting S/WOTs from a large actigraphy dataset.  Originally developed for children's actigraphy data collected by hip-worn ActiGraph sensors, this algorithm was seamlessly applied to adults' actigraphy data collected by wrist-worn Actiwatch sensors. Validated against the self-reported S/WOT event markers, our proposed algorithm showed good consistency with the markers, resulting in a small bias of less than 6 minutes in comparison with the markers. Using variance component models, we found that the between-subject variability and day-to-day variability within subjects contributed by far the largest percentage to the total variances in S/WOTs. This highlights the necessity of SW cycle detection when analyzing actigraphy data. While the between-subject variation can be partially modeled by certain covariates (e.g. age), individual habitual rhythms are largely random and cannot be learned well via population statistics. The large within-subject variation suggests that reporting average S/WOTs will lose considerable information about an individual's sleep patterns. \par

Our proposed algorithm has several advantages. First, this top-down approach converts the SW cycle detection problem into a change-point detection problem within bounded regions, which are the estimated circadian cycles. In contrast to previous methods that focused on supervised learning methods, the assumptions of CircaCP focus on the commonality of sleep patterns in humans. Following these assumptions, CircaCP captures the circadian cycles via a cosinor model and the day-to-day variability via CP detection bounded by each circadian cycle. Therefore, this method is directly transferable to other types of actigraphy, given apparent circadian rhythms and reasonably identified data distributions. For example, activity counts from Actiwatch and ActiGraph, although based on different proprietary algorithms, are continuous random variables with long-tail distributions, hence can both be modeled well by Gamma distributions where the dispersion (i.e. the inverse of the shape parameter) is explicitly estimated. For activity count data that are discrete and generated from a counting process (e.g. activity counts measured by infrared room occupancy sensors, or step counts from FitBit\textsuperscript{\textregistered} sensors), change-point detection algorithms based on Poisson distributions may achieve better detection accuracy. \par

Secondly, the only parameter that needs to be chosen in advance is the threshold to dichotomize the cosinor curve. Although we used a set of priors to initialize the cosinor model, those values are related to the circadian rhythms of human beings, thus can be applied to other datasets without tuning. Since the proposed algorithm circumvents the laborious collection of sleep labels and training classifiers for various actigraphy datasets collected by various devices with different configurations, this automated approach reduces the overall overhead in developing SW cycle detection algorithms. \par

Lastly, we elected to use the less computationally-demanding cosinor model instead of the Hill-transformed cosinor model in our previous publication (\cite{Chen2019}), so that the curve fitting step was more robust and much faster. Moreover, we improved the previous algorithm by a second round CP searching procedure, so that the mis-identified CPs could be better located in the second round. Since all of these steps can be done in linear time ($\mathcal{O}(n)$), with the algorithms implemented in Matlab 2021a using Intel i7 4-core CPU at 3.4GHz with 16.0GB RAM, it takes less than 15 minutes to complete the SW cycle detection on 1930 subjects' actigraphy data. \par

One limitation of using actigraphy for S/WOT detection is that the S/WOTs are effectively time-to-bed and time-to-get-up, whereas the precise time when the body physiologically and/or cognitively switches off is unattainable. For subjects who have afterhours sedentary behavior, actigraphy-based SW detection can underestimate their SOTs by a large margin (e.g. 7 pm can be identified as the SOT if the subject stays in bed after dinner until falling asleep). Our finding that the bias in SOT detected by CircaCP is behind SOT reported by markers partially confirmed this underestimation. Still, the framework of CircaCP (i.e. identifying CPs within a circadian cycle) can also be applied to synchronous heart rate data for even more precise S/WOT estimation.  Another limitation is that for certain subjects with severely interrupted circadian rhythms (mostly subjects who worked night shifts), the detection results contained large errors and were identified by the error detection method. Although this is a small percentage (less than 2\%) of the cohort, their poor regularity may be of greater interest in studies investigating the impact of occupation on health. For such actigraphy data, algorithms that take into account the characteristics of long sequences can be used (e.g. Gamma distribution-based Hidden Markov Models). Our future work will focus on fusing information from different sensor modalities and other types of estimation method. \par

The applications of our proposed algorithm are wide. In essence, it provides a unified framework for accurately detecting SW cycles and S/WOTs. In sleep research, multiple sleep-related metrics, such as sleep duration and S/WOT variability, can be derived based on accurate S/WOTs. For actigraphy analysis in general, CircaCP can also serve as a pre-processing step by segmenting the long sequences of actigraphy into two regions, within which time-series are relatively stationary, and thus time series models can be validly applied and metrics related to physical activity profiles can be accurately extrapolated. \par


%
%

\section{Financial disclosure}
The authors had no financial arrangements or connections to disclose. 

\section{Non-financial disclosure}
The authors have no potential conflicts of interest. 

\section{Author contributions statement}
S.C. conceived and conducted the experiments, analyzed and interpreted the results, wrote and revised the manuscript. X.S. contributed to data analysis and manuscript writing.

\section{Acknowledgments}
The Multi-Ethnic Study of Atherosclerosis (MESA) Sleep Ancillary study was funded by NIH-NHLBI Association of Sleep Disorders with Cardiovascular Health Across Ethnic Groups (RO1 HL098433).The National Sleep Research Resource was supported by the National Heart, Lung, and Blood Institute (R24 HL114473, 75N92019R002).


\bibliographystyle{biorefs}
\bibliography{SleepGeneric_Chen} 


\begin{figure}[h]
		\centering
		\includegraphics[scale=0.7]{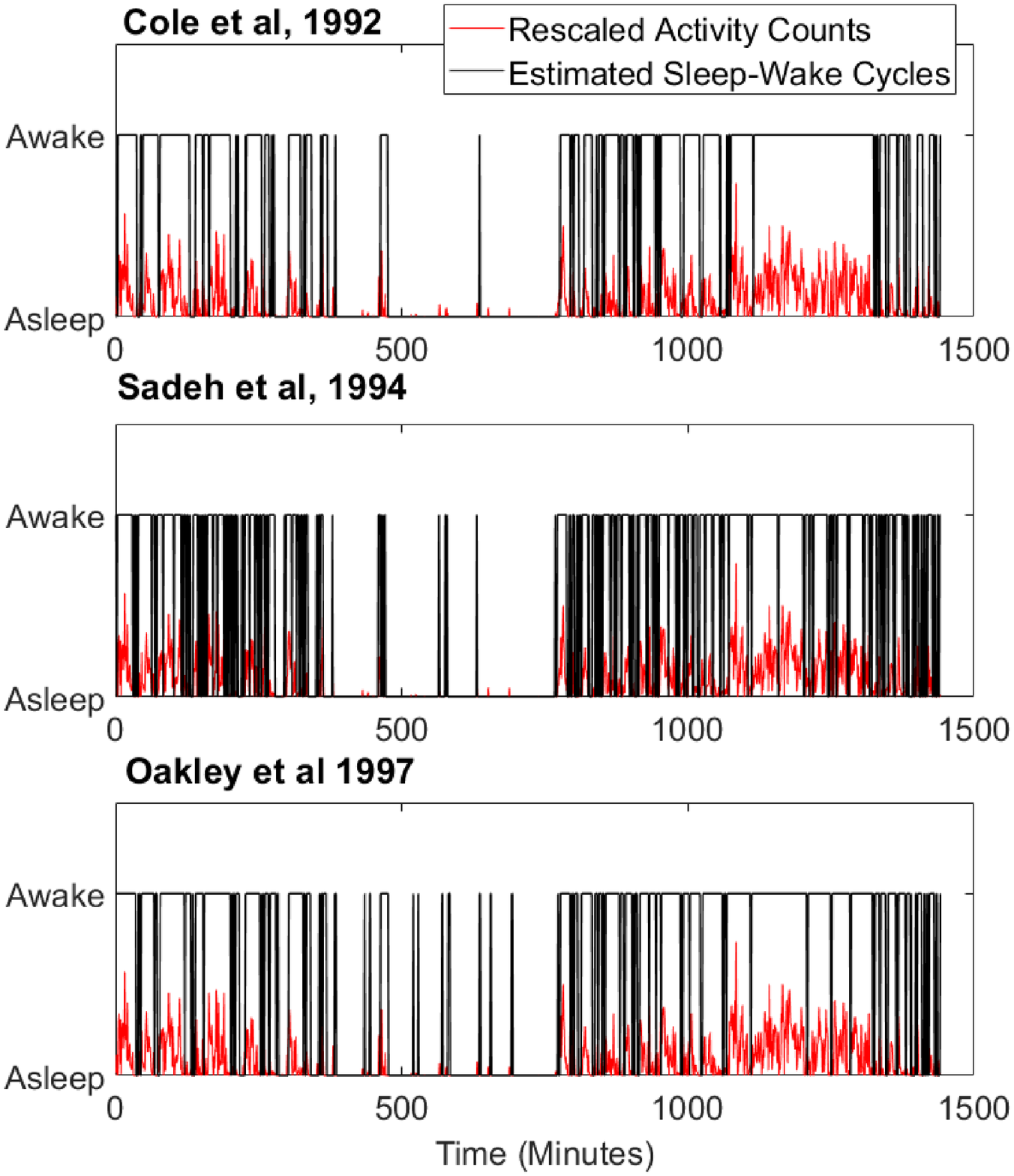}
	\caption{Sleep-wake states detected by three popular algorithms for one subject's actigraphy data from the MESA Sleep study}
\end{figure}

\begin{figure}[h!]
		\centering
		\includegraphics[scale=0.1]{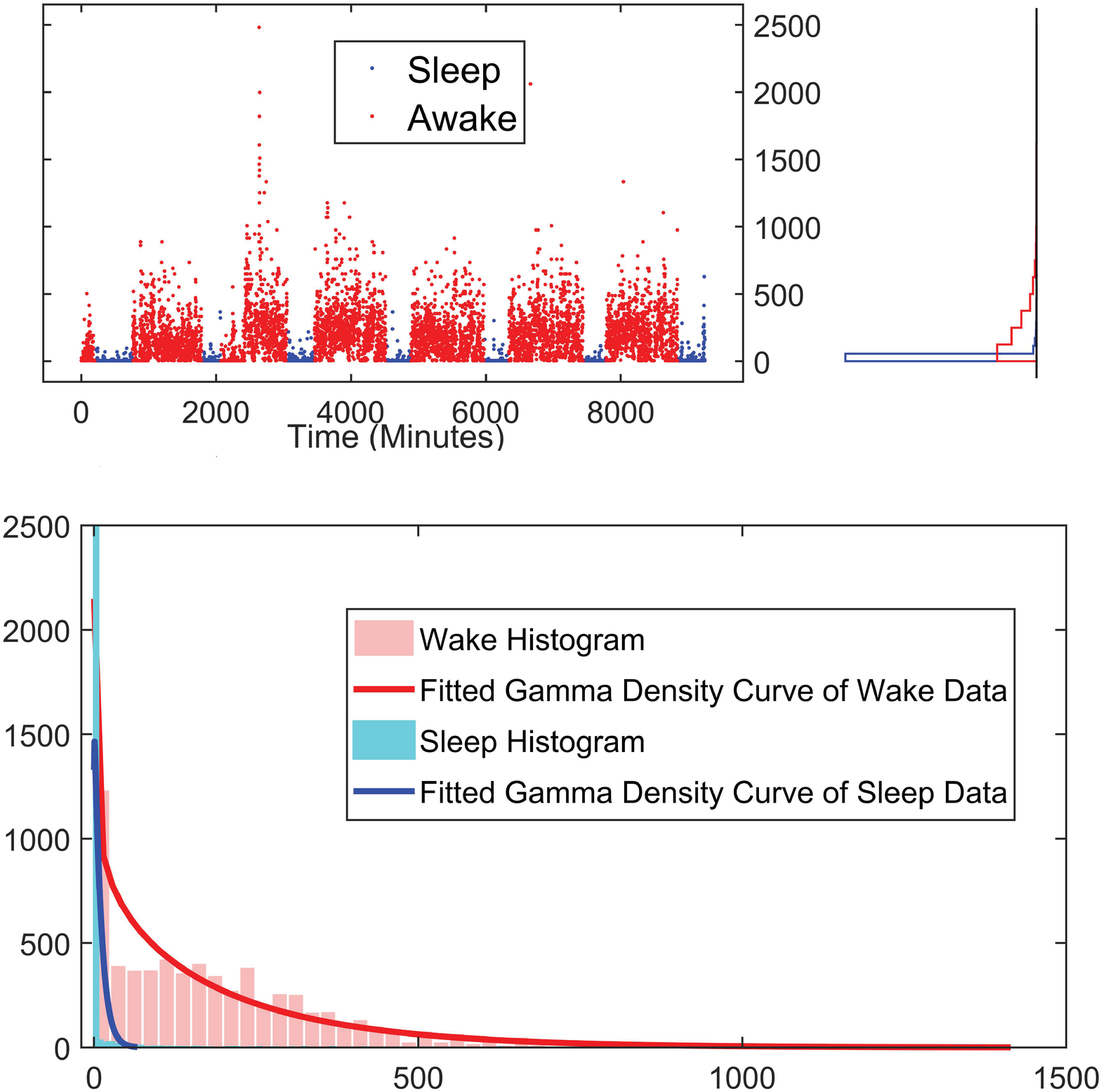}
	\caption{Histograms of actigraphy data by sleep period and awake periods. Actigraphy data within each period were fit by Gamma distributions.}
\end{figure}

\begin{figure}[h!]
		\centering
	\includegraphics[scale=0.75]{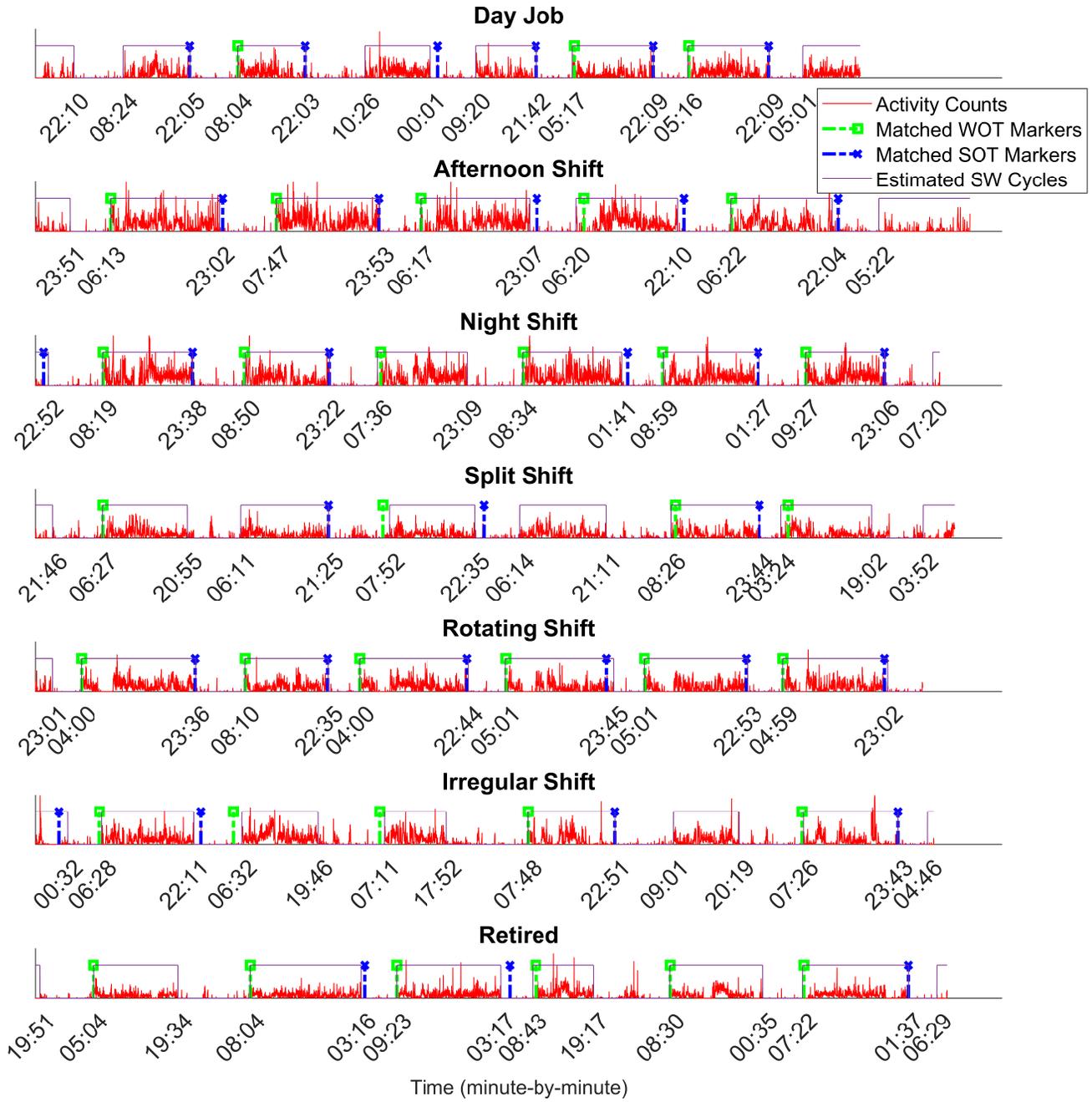}
	\caption{Visualization of SW detection results}
\end{figure}

\begin{figure}[h!]
		\centering
		\includegraphics[scale=0.6]{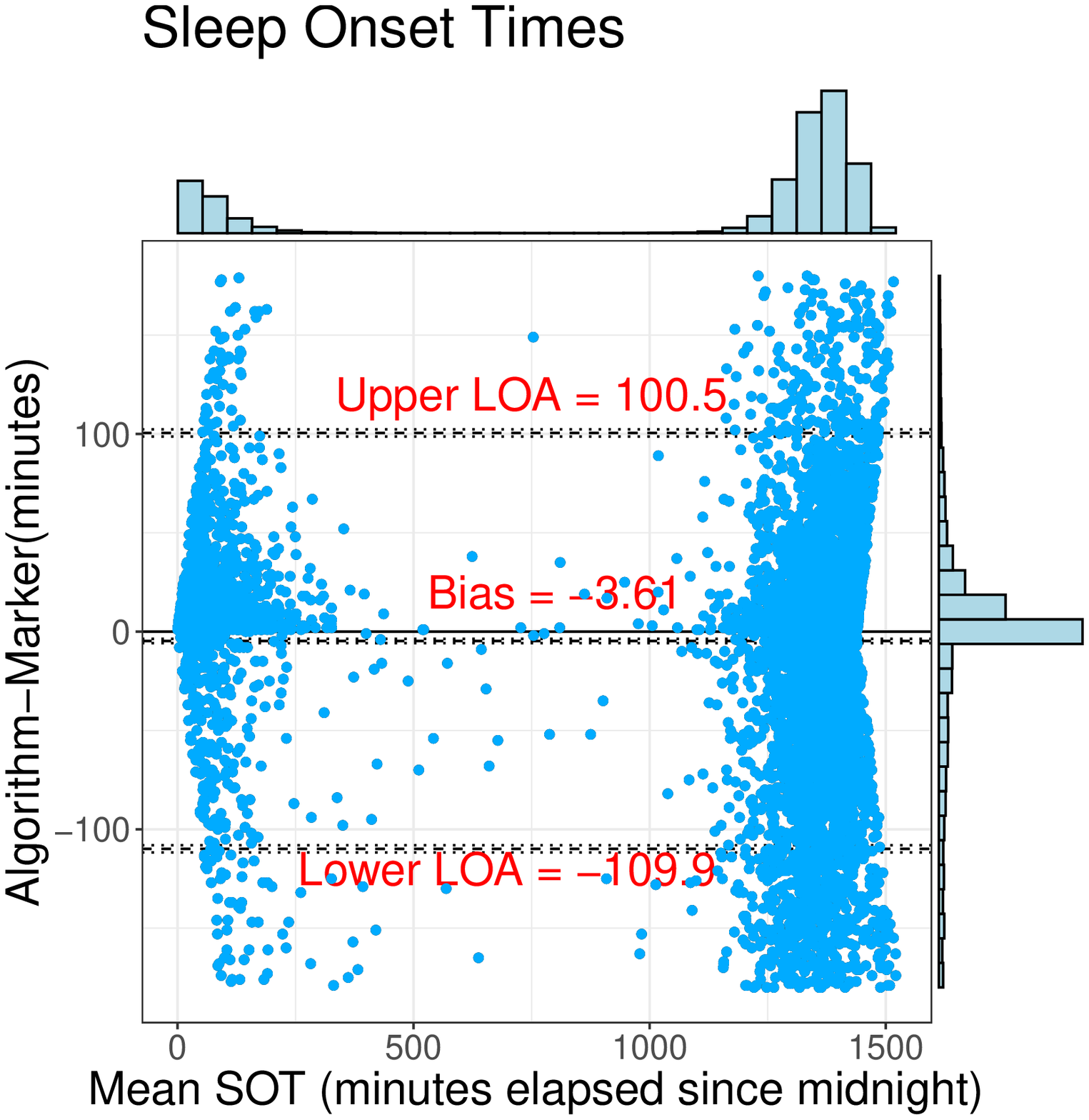}
	\caption{Bland-Altman plot for SOTs. The horizontal histogram shows the distribution of SOTs and the vertical histogram on the right is the error distribution. LOA indicates limit of agreement. }
\end{figure}

\begin{figure}[h!]
		\centering
		\includegraphics[scale=0.6]{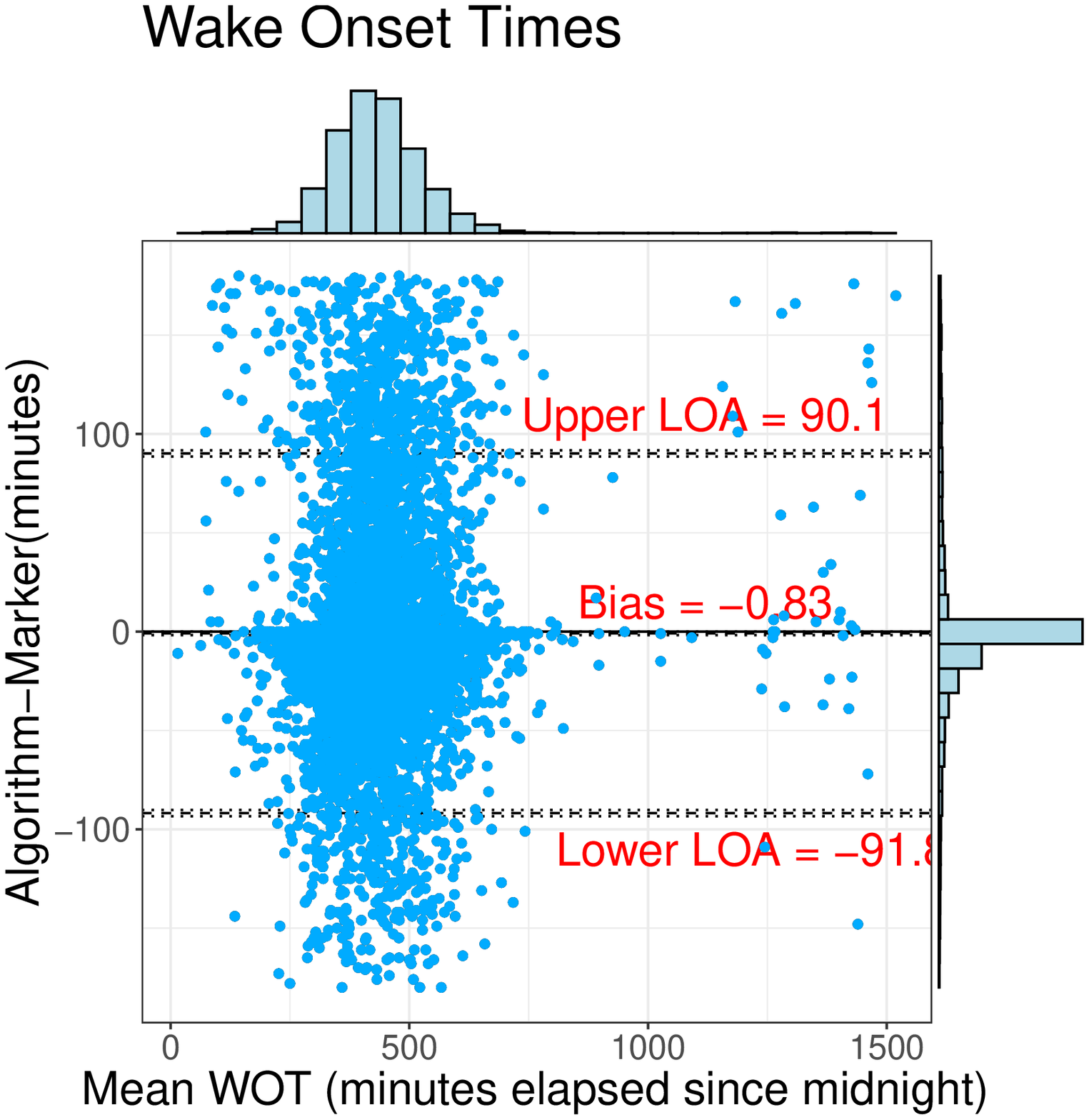}
	\caption{Bland-Altman plot for WOTs. The horizontal histogram shows the distribution of WOTs and the vertical histogram on the right is the error distribution. LOA indicates limit of agreement.}
\end{figure}

\newpage
\begin{table}[!h]
	\caption{Summary of Actigraphy Data in MESA Sleep Study.\label{tab1}}
    \footnotesize
    	\resizebox{\textwidth}{!}{
	\begin{tabular*}{\textwidth}{@{\extracolsep{\fill}}lllllll@{\extracolsep{\fill}}} 
		\toprule%
		& \multicolumn{6}{l}{\textbf{Actigraphy Quality Group}}                                             \\
		
		& G2  & G3   & G4     & G5   & G6  & G7 \\
		\midrule
		Original No.         & 692        & 573            & 49            &132           & 580          & 99 \\                                             
		No. for SW detection         & 627         & 525            & 46            &121           & 544          & 94 \\
		No. with poor SW results     & 15		   & 13		        & 0		        &2			   & 3			  &	1  \\
		No. with valid SW results     & 612       & 512             & 46               & 119     & 541             & 93   \\
		\midrule
		Included for Validation  & N=558  & N= 503    & N=45         & N=119       & N=539        & N=93   \\
		\textbf{Age}     &              &              &             &             &              &             \\
		Mean (SD)        & 71.2 (9.22)  & 69.7 (9.36)  & 68.2 (8.24) & 69.6 (9.19) & 68.5 (8.50)  & 67.7 (8.48) \\
		Median {[}Min, Max{]} & 71.0 {[}55.0, 94.0{]} & 69.0 {[}54.0, 94.0{]} & 68.0 {[}55.0, 86.0{]} & 68.0 {[}55.0, 92.0{]} & 67.0 {[}55.0, 91.0{]} & 66.0 {[}55.0, 88.0{]} \\
		\textbf{Gender}  &              &              &             &             &              &    \\
		Female           & 296 (53.0\%) & 292 (58.1\%) & 26 (57.8\%) & 53 (44.5\%) & 280 (51.9\%) & 49 (52.7\%) \\
		Male             & 262 (47.0\%) & 211 (41.9\%) & 19 (42.2\%) & 66 (55.5\%) & 259 (48.1\%) & 44 (47.3\%) \\
		\textbf{Race}    &              &              &             &             &              &             \\
		White Caucasian   & 127 (22.8\%) & 177 (35.2\%) & 20 (44.4\%) & 49 (41.2\%) & 287 (53.2\%) & 62 (66.7\%) \\
		Chinese American & 58 (10.4\%)   & 50 (9.9\%)   & 6 (13.3\%)  & 16 (13.4\%) & 54 (10.0\%)  & 6 (6.5\%)   \\
		African American  & 237 (42.5\%)  & 155 (30.8\%)  & 6 (13.3\%)  & 23 (19.3\%)  & 94 (17.4\%) & 14 (15.1\%)      \\
		Hispanic         & 136 (24.4\%) & 121 (24.1\%) & 13 (28.9\%) & 31 (26.1\%) & 104 (19.3\%) & 11 (11.8\%) \\
		\textbf{Work Schedule}    &              &              &             &             &              &             \\
		Day Shift         & 122 (21.9\%) & 156 (31.0\%) & 18 (40.0\%) & 41 (34.5\%) & 188 (34.9\%) & 37 (39.8\%) \\
		Afternoon Shift   & 25 (4.5\%)   & 7 (1.4\%)    & 0 (0\%)     & 3 (2.5\%)   & 12 (2.2\%)   & 0 (0\%)     \\
		Irregular Shift   & 25 (4.5\%)   & 36 (7.2\%)   & 3 (6.7\%)   & 4 (3.4\%)   & 33 (6.1\%)   & 12 (12.9\%) \\
		Night Shift       & 8 (1.4\%)    & 6 (1.2\%)    & 0 (0\%)     & 1 (0.8\%)   & 9 (1.7\%)    & 1 (1.1\%)   \\
		Split Shift       & 8 (1.4\%)    & 6 (1.2\%)    & 1 (2.2\%)   & 0 (0\%)     & 6 (1.1\%)    & 0 (0\%)     \\
		Rotating Shift    & 6 (1.1\%)    & 6 (1.2\%)    & 0 (0\%)     & 0 (0\%)     & 2 (0.4\%)    & 2 (0.4\%)   \\
		Retired          & 364 (65.2\%) & 286 (56.9\%) & 23 (51.1\%) & 70 (58.8\%) & 289 (53.6\%) & 41 (44.1\%) \\
      \bottomrule

	\end{tabular*}}

\end{table}

\begin{table}
	\caption{Results from linear mixed-effects models}\label{mixed}
	\footnotesize
	\resizebox{\textwidth}{!}{
	\begin{tabular}{l|ccc|ccc}
		\toprule
		\textbf{}        & \multicolumn{3}{c|}{\textbf{SOT}}          & \multicolumn{3}{c}{\textbf{WOT}}           \\
		\textit{Predictors} & \textit{Estimates} & \textit{CI}       & \textit{p}      & \textit{Estimates} & \textit{CI}     & \textit{p}      \\
		\midrule
		(Intercept)     & 1406.3  & 1388.1 – 1424.6 & \textbf{\textless 0.001} & 404.7 & 386.6 – 422.7 & \textbf{\textless 0.001} \\
		G2               & -9.7 & -27.8 – 8.5  & 0.297            & 27.1  & 9.1 – 45.2    & \textbf{0.003}  \\
		G3               & -9.1  & -27.0 – 8.8  & 0.320           & 11.1  & -6.6 – 28.9   & 0.220            \\
		G4               & -13.8 & -42.6 – 15.1  & 0.350           & 4.1   & -24.8 – 33.0  & 0.780           \\
		G5               & -8.8  & -30.7 – 13.1 & 0.430            & 11.6  & -10.2 – 33.2   & 0.298            \\
		G6               & -13.8 & -31.5 – 3.9  & 0.126            & 2.4   & -15.1 – 19.9  & 0.789           \\
		Algorithm        & -4.7  & -6.5 – -2.9  & \textbf{\textless 0.001} & -0.80  & -3.0 – 1.4    & 0.459           \\
		Afternoon Shift  & 71.0   & 46.3 – 95.7     & \textbf{\textless 0.001} & 68.8 & 44.2 – 93.3   & \textbf{\textless 0.001} \\
		Irregular Shift  & 25.4  & 9.0 – 41.8   & \textbf{0.002}  & 31.1   & 14.6 – 47.6   & \textbf{\textless 0.001} \\
		Night Shift      & 39.7   & 7.1 – 72.3  & \textbf{0.017}  & 202.9 & 169.4 – 236.4 & \textbf{\textless 0.001} \\
		Retired          & 27.8  & 18.6 – 37.0  & \textbf{\textless 0.001} & 43.5  & 34.3 – 52.8   & \textbf{\textless 0.001} \\
		Rotating Shift   & -4.2   & -44.4 – 35.9 & 0.836           & 15.4  & -18.27 – 57.70  & 0.446           \\
		Split Shift      & 46.3  & 11.2 – 81.4  & \textbf{0.010}   & 52.3   & 16.6 – 87.9    & \textbf{0.004}  \\
		Age              & -1.4   & -1.9 – -1.0  & \textbf{\textless 0.001} & -0.5   & -1.0 – -0.0   & \textbf{0.030}  \\
		Male             & 0.7  & -6.8 – 8.1   & 0.859             & -1.2  & -8.6 – 6.3   & 0.758           \\
		Chinese American & 7.9   & -5.2 – 21.0  & 0.239           & 2.7   & -10.4 – 15.8  & 0.683           \\
		African American & 12.1  & 2.5 – 21.7   & \textbf{0.013}   & -0.1    & -9.7 – 9.5   & 0.98           \\
		Hispanic         & -6.1  & -16.1 – 3.9  & 0.231           & -11.6  & -21.6 – -1.6   & \textbf{0.023}      \\
    \bottomrule

	\end{tabular}}
\end{table}

\begin{table}
	\caption{Results of variance component analysis. All numbers are presented as percentages.}\label{varcomp}
	\centering
	\begin{tabular}{l|l|l}
		\toprule
		& SOT   & WOT  \\
		\midrule
		Algorithm    & 0.154  & 0.002 \\
		Subject      & 59.122 & 43.736  \\
		Subject/Day  & 33.675  & 47.599 \\
		Residual    & 7.049	& 8.663	\\
		\bottomrule
	\end{tabular}
\end{table}

\end{document}